\documentclass[authoryear,preprint,12pt]{elsarticle}
\pdfoutput=1 
\makeatletter
\def\ps@pprintTitle{%
	\let\@oddhead\@empty
	\let\@evenhead\@empty
	\let\@oddfoot\@empty
	\let\@evenfoot\@oddfoot
}
\makeatother

\usepackage{array} 
\usepackage{textcomp} 
\usepackage[breaklinks]{hyperref} 
\usepackage[table,xcdraw]{xcolor} 
\graphicspath{ {images/} } 
\usepackage[none]{hyphenat} 
\usepackage{siunitx} 
\usepackage{amssymb} 
\usepackage{subfig} 
\usepackage{amsmath} 
\usepackage{mhchem}

\begin{document} \sloppy 

\begin{frontmatter}

\title{ Spectral clustering tools applied to Ceres in preparation for OSIRIS-REx color imaging of asteroid (101955) Bennu}

\author[label1,label2]{Juan Luis Rizos} \ead{jlrizos@iac.es} 
\author[label1,label2]{Julia de Le\'on} 
\author[label1,label2]{Javier Licandro} 
\author[label3]{Humberto Campins} 
\author[label1,label2,label5]{Marcel Popescu} 
\author[label3]{Noem\'i Pinilla-Alonso} 
\author[label4]{Dathon Golish} 
\author[label3]{Mario de Pr\'a} 
\author[label4]{Dante Lauretta}

\address[label1]{Instituto de Astrof\'isica de Canarias, C/V\'ia L\'actea s/n, E-38205 La Laguna, Tenerife, Spain} 
\address[label2]{Departamento de Astrof\'isica, Universidad de La Laguna, E-38206 La Laguna, Tenerife, Spain} 
\address[label3]{Physics Department, University of Central Florida, P.O. box 162385, Orlando, FL 32816-2385, USA} 
\address[label4]{Lunar and Planetary Laboratory, University of Arizona, 1415 N. Sixth Ave. Tucson, AZ 85705-0500, USA}
\address[label5]{Astronomical Institute of the Romanian Academy, 5 Cu\c{t}itul de Argint, 040557 Bucharest, Romania}

\begin{abstract} The OSIRIS-REx asteroid sample-return mission is investigating primitive near-Earth asteroid (101955) Bennu. Thousands of images will be acquired by the MapCam instrument onboard the spacecraft, an imager with four color filters based on the Eight-Color Asteroid Survey (ECAS): $b$' (473 nm), $v$ (550 nm), $w$ (698 nm), and $x$ (847 nm). This set of filters will allow identification and characterization of the absorption band centered at 700 nm and associated with hydrated silicates. In this work, we present and validate a spectral clustering methodology for application to the upcoming MapCam images of the 
surface of Bennu. Our procedure starts with the projection, calibration, and photometric correction of the images. In a second step, we apply a K-means algorithm and we use the Elbow criterion to identify natural clusters. This methodology allows us to find distinct areas with spectral similarities, which are characterized by parameters such as the spectral slope $S$' and the center and depth of the 700-nm absorption band, if present. We validate this methodology using images of (1) Ceres from NASA's Dawn mission. In particular, we analyze the Occator crater and Ahuna Mons. We identify one spectral cluster--located in the outer parts of the Occator crater interior--showing the 700-nm hydration band centered at 698 $\pm$ 7 nm and with a depth of 3.4 $\pm$ 1.0 \%. We interpret this finding in the context of the crater's near-surface geology.

\end{abstract}


\end{frontmatter} 


\section{Introduction} \label{section1}

Primitive asteroids are residual planetary building blocks. They contain solar nebula materials and pre-solar grains. Primitive (C-complex) asteroids have low-albedo surfaces ($p_V < 15$ \%) and almost featureless spectra. These asteroids are likely composed of carbon compounds, organics and water-bearing 
minerals, as their spectra resemble those of the carbonaceous chondrites. Because these asteroids have hydrated minerals and organic compounds, they are of particular interest as they could answer fundamental questions about how water arrived on Earth. Also, because of the composition and accessibility of primitive 
near-Earth asteroids (NEAs), there are of growing interest in them for in situ resource utilization (ISRU). The cost of extracting fuel and building materials from asteroids may soon be lower than the cost of launching these resources from Earth, thus creating the foundation for a self-sustaining space economy. Two sample-return missions, JAXA's Hayabusa2 and NASA's OSIRIS-REx (Origins, Spectral Interpretation, Resource Identification, and Security--Regolith Explorer), are visiting two primitive NEAs, (162173) Ryugu and (101955) Bennu, respectively. Among the goals of both missions is the identification of resources for future 
exploitation of asteroids.

OSIRIS-REx was successfully launched on September 8, 2016 and has recently encountered NEA Bennu. Discovered on September 11, 1999 \citep{Ticha1999}, Bennu is a B-type carbonaceous asteroid \citep{Clark2011} classified as one of the most potentially hazardous NEAs \citep{Chelsey2014}. The spacecraft is equipped with different scientific instruments to measure Bennu's physical, geological, and chemical properties, and select an area on the surface of the asteroid to collect a sample. Specially designed for this purpose is the OSIRIS-REx Camera Suite (OCAMS), a set of three cameras (PolyCam, SamCam, and MapCam) to image Bennu. We focus on an analysis tool for the color data that will be obtained by MapCam.

MapCam is a medium-field imager, with a 125 mm focal length f/3.3 optical system that will provide a $\sim$70 mrad ($4 ^{\circ}$) field of view \citep{Rizk2018}. This camera will produce the images needed for the development of color maps and global shape models. To this purpose, the instrument contains four 
filters with passbands based on the Eight-Color Asteroid Survey (ECAS) filter passbands \citep{Tedesco1982}: $b'$, $v$, $w$, and $x$, centered at 473, 550, 698, and 847 nm respectively \citep{Lauretta2017}. The original ECAS $b$ filter has been shifted toward longer wavelengths to improve optical and radiometric performance and to minimize aging effects due to radiation \citep{Rizk2018}. MapCam will provide color information with very high spatial resolution. This set of filters will permit the detection and characterization of an absorption band centered at 700 nm, if present. This band is associated with the presence of phyllosilicates---i.e., silicates that have been altered by liquid water---and strongly correlates with the well-known absorption feature at 3 $\mu$m indicative of hydration \citep{Vilas1994,Rivkin2015}. These filters will also measure the spectral slope, which can be used to identify the effects of space weathering. Global and site-specific color ratio maps will help identify areas of interest (at a resolution $\lesssim$ 1 cm) and select a sampling site.

In preparation for the scientific exploitation of the MapCam data, we have analyzed the images of (1) Ceres obtained by the NASA Dawn mission and its Framing Camera (FC). Dwarf planet Ceres, although substantially larger than asteroid Bennu, is also a primitive body: both local exposed H$_{2}$O and hydrated minerals have been found on its surface \citep{Combe2016,Ammannito2016,Nathues2017}. The FC is equipped with a set of seven color filters that cover a similar wavelength region as the MapCam filters. We analyze the color images of the surface of Ceres obtained by the FC and develop a robust 
methodology than we can apply to the color images of the surface of Bennu taken by MapCam.

Our approach was to select images from (1) Ceres showing the different features that we expect to find on Bennu (e.g., boulders, dark and bright regions and craters). We downloaded the Ceres images from the Small Bodies Node of the Planetary Data System (PDS\footnote{https://pds.nasa.gov/}). In Section 
\ref{Section2} we describe the different calibration processes that we followed to prepare the images for analysis. In Section \ref{Section3} we explain how we analyzed the data using spectral clustering techniques, which identified regions (clusters) with spectral similarities, and how we characterized these clusters. 
Finally, the results are presented in Section \ref{Section4} and discussed in Section \ref{Section5}.


\section{Data Preparation} \label{Section2}

\begin{figure}[h!] 
\centering 
\includegraphics[width=7cm, height=8cm]{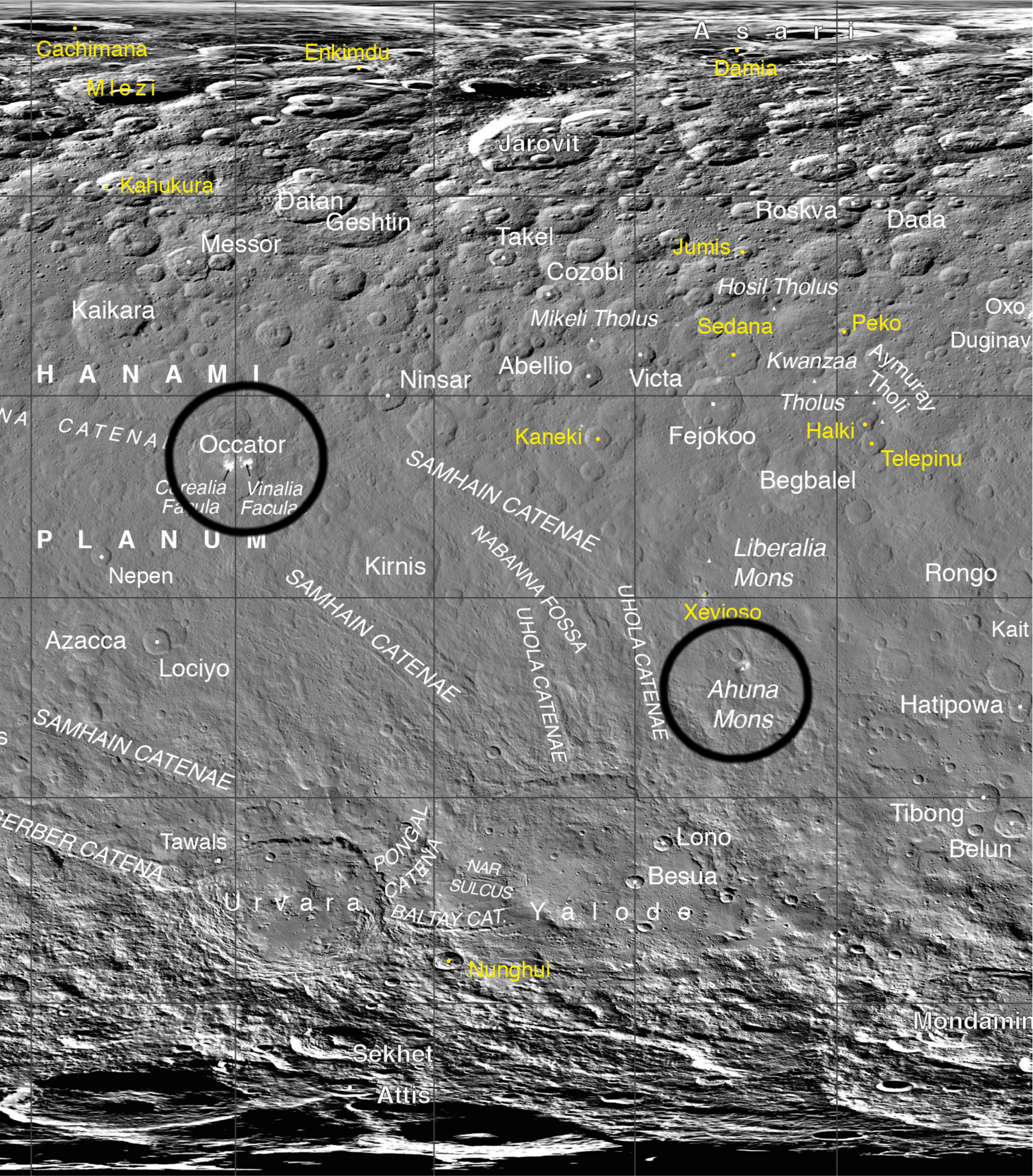} 
\caption{Cylindrical projection of (1) Ceres with longitude range between 210$^{\circ}$ and 30$^{\circ}$ East and complete latitude range. Black circles mark the regions analyzed in this paper. More information about the selected regions is shown in Table \ref{table: 1}. Credits: NASA/JPL-Caltech/UCLA/MPS/DLR/IDA (Thomas Roatsch).} \label{fig: ceres} \end{figure}

The NASA Dawn spacecraft orbited dwarf planet (1) Ceres, acquiring tens of thousands of images from different altitudes. The FC is an onboard instrument equipped with a charge coupled device (CCD) providing a field of view of ${5.5}^{\circ}$x${5.5}^{\circ}$, one clear filter (F1), and seven color filters (F2 to F8) covering the wavelengths from the visible to the near-infrared \citep{Sierks2011}. We used raw and calibrated images of Ceres obtained with the FC. We selected 
images of two regions on the surface of Ceres, marked with black circles in Fig. \ref{fig: ceres} and listed in Table \ref{table: 1}. Our primary selection criterion was to have different spectral and albedo characteristics to test the robustness and suitability of our method. The regions are the Occator crater, where bright spots suggest the existence of salts \citep{Nathues2015} and Ahuna Mons, also listed among the bright areas identified on the surface of Ceres.

\begin{table}[h!] 
\begin{center} 
\footnotesize 
\setlength{\tabcolsep}{6pt} 
\begin{tabular}{lcccc} 
\hline 
Region & Images & Resolution & Center Latitude & Center Longitude \\ 
& ID & (m/pixel) & ($^{\circ}$) & ($^{\circ}$)\\ 
\hline 
Occator Crater & 40745--40751 & 139.4 & 19.82 & 239.3 \\ 
Ahuna Mons & 45072--45078 & 137.4 & -10.48 & 316.2 \\ 
\hline 
\end{tabular} 
\end{center} 
\caption{Selected regions of (1) Ceres analyzed in this work. The regions are marked with black circles in Fig. \ref{fig: ceres}.} 
\label{table: 1} 
\end{table}

The first step was to download the images from the PDS calibrated in spectral radiance, $L$, and to convert them to radiance factor using the following equation: 

\begin{equation} 
\label{equation_new}
RADF_{i}=\frac{\pi d^{2}L_{i}}{F_{i}} 
\end{equation} 
where $d$ is the distance of Ceres to the Sun at the time of the observation (in meters), and $F_{i}$ the effective solar flux for the $i$-th color filter\footnote{The 
effective solar flux values used in this work are published in the documentation prepared by S. Sch{\"o}rder \& P. Gutierrez-Marques about the FC, available in the PDS Small Bodies Node (https://sbnarchive.psi.edu/pds3/dawn/fc/DWNCAFC2\_1B/DOCUMENT/CALIB\_PIPELINE)}. The radiance factor $RADF$ is equivalent to the commonly used notation $I/F$ \citep{Li2015}, also called reflectance, a dimensionless magnitude defined as the ratio of the bidirectional reflectance of a surface to that of a perfectly scattering surface illuminated from the normal direction. 

\begin{figure}[h!] 
\centering 
\includegraphics[width=\textwidth]{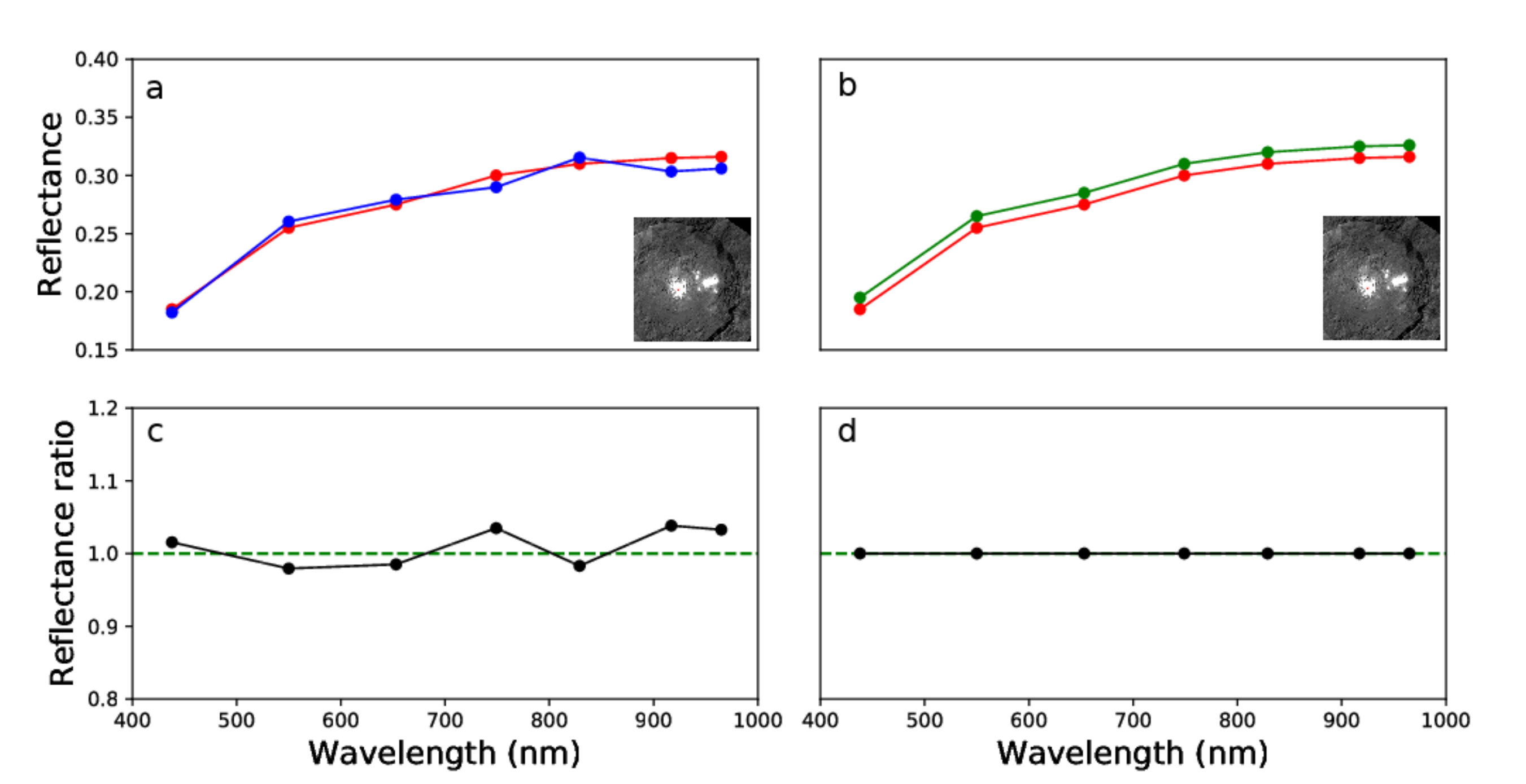} 
\caption{ Comparison between our computed reflectances and those published in the literature. a) Reflectances in the Occator crater (Cerealia facula) computed using the calibrated PDS images (blue), and the corresponding reflectances (red) computed by \cite{Schroder2017}, for the area enclosed by the red square in 
Fig. 13a of that paper. b) Same as a) but after applying our calibration to the images (green). Reflectances obtained by \cite{Schroder2017} are shown in red and with an offset for a better comparison. Panels c) and d) show the ratio between the reflectances in a) and b), respectively.} \label{fig:comparison} 
\end{figure}

However, we noticed that the obtained reflectances presented inconsistencies when compared with published data. To illustrate this point, we downloaded from the PDS the same calibrated images used by \cite{Schroder2017} to compute the reflectances inside a small square region located in the Cerealia facula of the Occator crater (see Fig. 13a in \citealt{Schroder2017}). Our reflectances computed for calibrated PDS images are shown in blue in Fig. \ref{fig:comparison}a, whereas the reflectances calculated by \cite{Schroder2017} are shown in red. The observed differences are better visualized using the ratio of the reflectances (Fig. \ref{fig:comparison}c). Although these differences are within the error, they systematically appeared when using the calibrated PDS images, independently of the region selected to compute the reflectances. Therefore, we decided to use raw images and to calibrate them ourselves.

Our calibrations are based on the pipeline used by \cite{Schroder2017} and described in detail in \citet{Schroder2013,2014Schro}. The first step was to take raw 
images obtained using the $i$-th color filter and to calibrate them in spectral radiance, $L_{i}$ (W$^{-2}$nm$^{-1}$sr$^{-1}$), using the following equation: \begin{equation} 
\label{equiation1} 
L_i=\frac{(A_{i}-S-b)/t_{exp}-D-I_{i}}{R_{i}N_{i}} 
\end{equation} 
where $A_{i}$ is the raw image in the $i$-th color filter; $S$ is the smear, a charge gradient from image top to bottom due to exposure during the reading process; $b$ is the bias value; $t_{exp}$ is the exposure time; $D$ is the dark current image; $I_{i}$ is the in-field stray light image, an effect caused by the diffraction of the incident light in the CCD and its consequent reflection in the filters; $R_{i}$ is the responsivity factor, a measure of how light is transmitted in each filter; and 
$N_{i}$ is the normalized flat field\footnote{The necessary files to apply this step, like the flats or stray light patterns, are available via the Dawn Data website 
(http://dawndata.igpp.ucla.edu/).}. Once we calibrated the images in spectral radiance, the next step was to calibrate them in radiance factor, $RADF$, using equation \ref{equation_new}. The reflectances obtained using our calibration procedure in the case of the example area inside the Occator crater are shown in Fig. \ref{fig:comparison}b (green). These results are in excellent agreement with the values obtained by \cite{Schroder2017} (red). This agreement is better visualized using the ratio of the two reflectances shown in Fig. \ref{fig:comparison}d.

After were calibrated the images, the second step was to project them ---that is, to systematically transform the latitudes and longitudes of locations from the surface into positions on a plane. We used the Integrated Software for Imagers and Spectrometers (ISIS) to project the images.\footnote{ISIS is a robust tool to manipulate imagery collected by NASA planetary missions. It allows us to analyze three-dimensional cubes collected from imaging spectrometers, to project images, and to create mosaics (https://isis.astrogeology.usgs.gov).} We applied an equirectangular projection, also called the equidistant cylindrical projection. Once the images were calibrated and projected, the final step was to correct them photometrically.

\subsection{Photometric corrections} \label{subsection2}

\begin{figure}[!ht] 
\centering 
\includegraphics[width=0.48\textwidth]{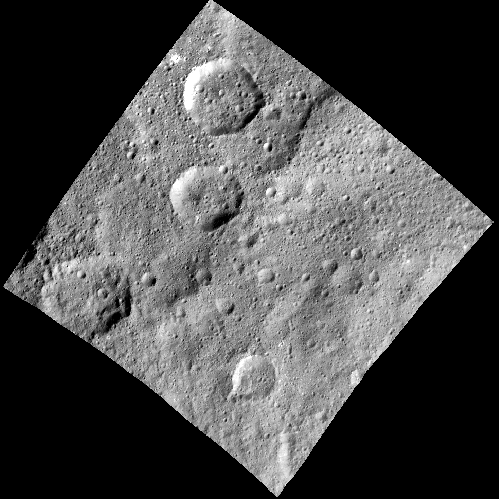} 
\includegraphics[width=0.48\textwidth]{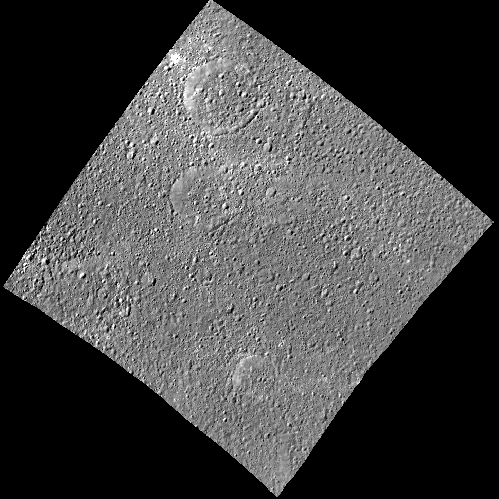} 
\caption{Example of the application of the photometric correction. Shown are equirectangular projected images of the south of Kaneki crater observed using filter F8 (438 nm) without (left) and with (right) photometric correction. We used the coefficients values shown in Table \ref{table: 2} for the phase function and the parameter-less Akimov disk function.} 
\label{fig: correction} 
\end{figure}

The last step in the calibration process, and maybe the most critical one, is the photometric correction. To accurately interpret the features detected in the surface of Ceres in terms of actual differences in the material, it is necessary to correct the images for differences in brightness caused by different viewing geometries. To do this, we calculate the reflectance value of each pixel for a set of reference values of incidence, emission, and phase angles ($i_{0},e_{0},\alpha_{0}$). To obtain these angles we use the ISIS software and a Digital Terrain Model (DTM) based on the Dawn High Altitude Mapping orbit images and derived  using the stereo photogrammetry (SPG) method. The DTM, developed by the Dawn team, has a lateral spacing of $\sim$136.7 m/pixel (60 pixel/degree) and 
is available at the PDS website. Once we make the corrections, each pixel has the reflectance that it would have if it had been observed from these reference angles. In this way, the absolute value of the reflectance changes in such a way that local topography almost disappears and intrinsic reflectivity variations over the surface become more evident (an example is shown in Fig. \ref{fig: correction}). 

There are different photometric models that reproduce the behavior of the scattered light on a surface. To correct the images, we computed the reflectance in each pixel of the image using this equation:

\begin{equation}
\label{eq:1} 
r(i_{0},e_{0},\alpha_{0})=\frac{r(i_{m},e_{m},\alpha_{m})}{r_{model}(i_{m},e_{m},\alpha_{m})}\;r_{model}(i_{0},e_{0},\alpha_{0}) 
\end{equation} 
where $r(i_{0},e_{0},\alpha_{0})$ is the photometrically corrected reflectance and $i_{m}$, $e_{m}$, and $\alpha_{m}$ are the incidence, emission, and phase angles corresponding to the scattering geometry of the measured reflectance. The most commonly used reference geometry is $(i_{0},e_{0},\alpha_{0})=({30}^{\circ},{0}^{\circ},{30}^{\circ})$. We used this reference set of angles in this work, which corresponds to $(\beta_{0},\gamma_{0},\alpha_{0})=({0}^{\circ},{0}^{\circ},{30}^{\circ})$ where $\beta$ and $\gamma$ are the photometric latitude and longitude respectively \citep{Li2015}. These coordinates are defined by the position of the Sun and the observer and are related to the previous coordinates as follows: $$ \gamma=\arctan\frac{\cos(i)-\cos(e)\cos(\alpha)}{\cos(e)\sin(\alpha)} $$ $$ \beta=\arccos\frac{\cos(e)}{\cos(\gamma)} $$

\begin{figure}[!h] 
\centering 
\includegraphics[width=\textwidth]{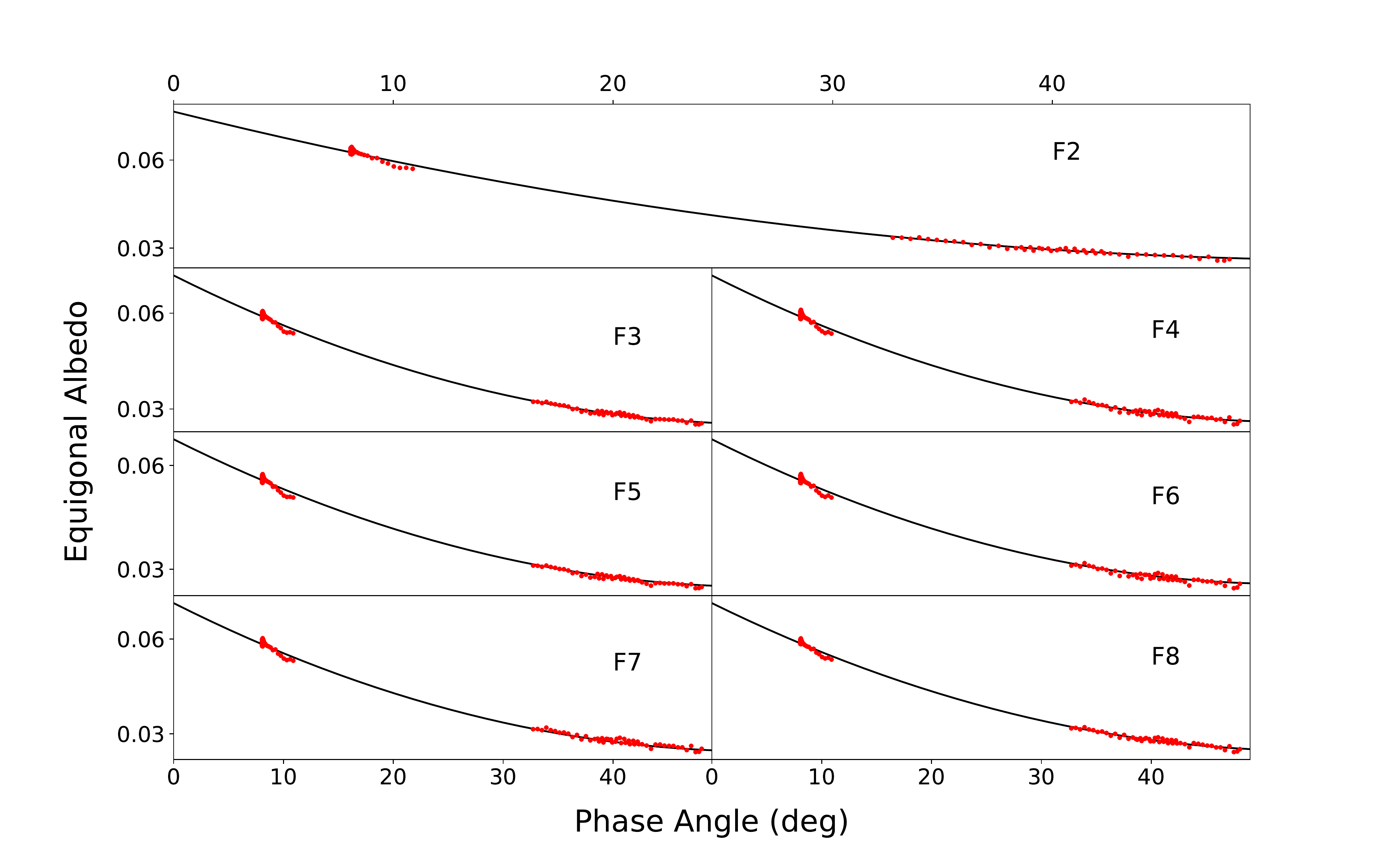} 
\caption{Polynomial phase curves for the equigonal albedo in the eight FC color filters valid in combination with the parameter-less Akimov disk function. Red points are the average equigonal albedo of each of the 75 images that we used. Black curves are the second-order 
polynomial fits. The phase angle ranges from $7^{\circ}$ to $49^{\circ}$.} 
\label{fig: fits} 
\end{figure}

We used the parameter-less Akimov model, which was validated by \cite{Schroder2017} as one of the best models for low phase angle images of Ceres such as the ones that we are correcting here. This model describes the radiance factor $RADF$ as a product of a disk function $D$ and a phase function $A_{eq}$ (also known as the equigonal albedo). Given the equigonal albedo curves for each filter obtained from observational data (Fig. \ref{fig: fits}) as described below, we assumed that the phase function is a second-order polynomial:

\begin{equation} RADF=A_{eq}(\alpha)\;D(\alpha,\beta,\gamma) 
\end{equation}

\begin{equation} A_{eq}(\alpha)=\sum _{i=0}^{2}C_{i}\;\alpha^{i} 
\end{equation}

\begin{equation} 
D(\alpha,\beta,\gamma)=\cos(\frac{\alpha}{2})\;\cos(\frac{\pi}{\pi-\alpha}(\gamma-\alpha/2))\;\frac{\cos(\beta)^\frac{\alpha}{\pi-\alpha}}{\cos(\gamma)} 
\end{equation}

To apply the model to our images, we have to know the value of the polynomial coefficients of the phase function, $C_{i}$, which is found by fitting the observational data to our photometric model. We first calibrated raw images of Ceres from the first mapping orbit, the RC3 phase, having different phase angles and different rotation states. We used a total of 75 images for each color filter (IDs between 36531 and 37263), spanning phase angles between 7$^{\circ}$ and 49$^{\circ}$, and corresponding to the northern, equatorial, and southern regions observed in this stage of the mission. Then, we obtained the equigonal albedo of each pixel by dividing our measured reflectance by the disk function of the Akimov model. We only used those pixels having emission or incidence angles lower than ${80}^{\circ}$, because it is in this range where our photometric model has precise performance \citep{Schroder2017}. For each image, we then averaged the equigonal albedo and the phase angle of all the pixels, and ended with $75$ points (red points in Fig. \ref{fig: fits}). Finally, we fit our second-order 
polynomial phase function to these 75 points with a Levenberg-Marquardt algorithm (black curve in Fig. \ref{fig: fits}). In Table \ref{table: 2}, we present the values obtained for the polynomial coefficients. 

\begin{table}[!h] 
\begin{center} 
\small 
\setlength{\tabcolsep}{7pt} 
\begin{tabular}{lccc} 
\hline 
Filter & $C_{0}$ & $C_{1}$ & $C_{2}$ \\ 
\hline 
F2 (555 nm)& 0.0765 & - 0.00186 & 0.0000171 \\ 
F3 (749 nm)& 0.0718 & - 0.00172 & 0.0000159 \\ 
F4 (917 nm)& 0.0685 & - 0.00164 & 0.0000155 \\ 
F5 (965 nm)& 0.0676 & - 0.00159 & 0.0000148 \\ 
F6 (829 nm)& 0.0698 & - 0.00168 & 0.0000160 \\ 
F7 (653 nm)& 0.0714 & - 0.00175 & 0.0000163 \\ 
F8 (438 nm)& 0.0696 & - 0.00172 & 0.0000157 \\ 
\hline 
\end{tabular} 
\end{center} 
\caption{Coefficients for the polynomial phase functions for each color filter, obtained with the parameter-less Akimov model. These coefficients are valid for phase angles in the range 7$^{\circ}$ to 49$^{\circ}$ with emission and incidence angles lower than 80$^{\circ}$.} 
\label{table: 2} 
\end{table}

Once we obtained the parameter values for our photometric model, we applied the photometric corrections to the images calibrated in $RADF$ and showing the areas of interest, indicated in Table \ref{table: 1}.


\section{Methodology: clustering technique} \label{Section3}

\begin{figure}[!htbp] 
\centering 
\includegraphics[width=0.95\textwidth]{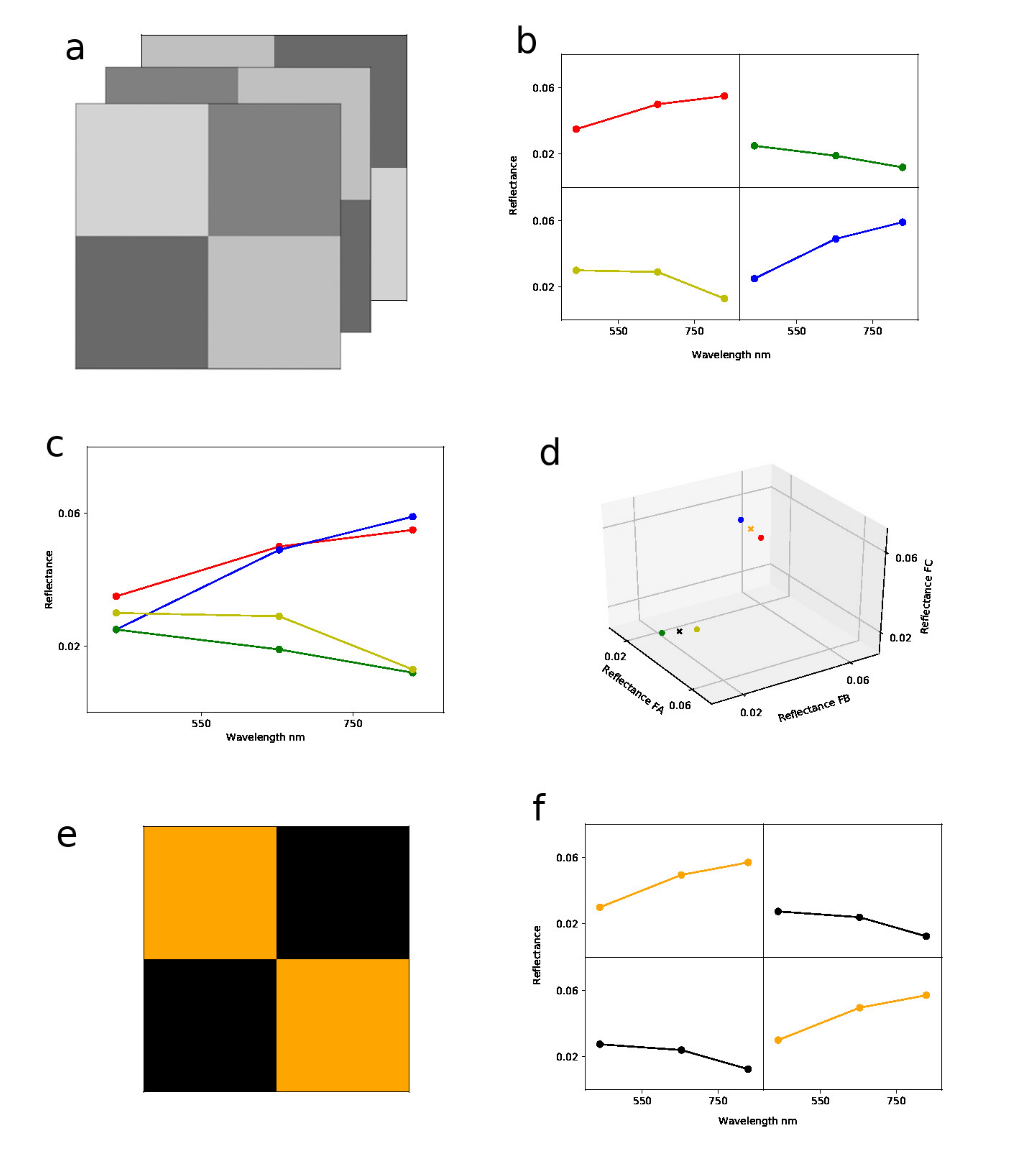} 
\caption{Illustration of the clustering technique applied to four-pixel images measured with three filters, A, B, and C, centered at 438, 653, and 829 nm respectively. The image is represented on a gray scale (a). The reflectance spectra of each pixel are represented separately (b) and in a standard plot for 
comparison (c). These spectra are represented in the three-dimensional spectral space (d). In this space, the x's correspond to the representative clusters of each subgroup. According to the distances to each of these two points, each pixel in the image is assigned to one group or the other (e). The last panel (f) shows the shape of each representative group.} 
\label{fig:kmeans} 
\end{figure}

With the images calibrated and photometrically corrected, we proceeded to analyze them using spectral clustering to identify zones on the surface of (1) Ceres with similar spectroscopic properties. This technique has been validated in previous works that applied it to different objects in the Solar System 
\citep{marzo2006,Pinilla-Alonso2011, julia2012}. We use the K-Means algorithm \citep{macqueen1967}, which is an unsupervised, partitioning\footnote{Partitioning or optimization techniques allow relocation of the spectral points during the clustering 
procedure, that is, poor initial partitions or groupings can be corrected later during the process.} technique \citep{Chirstopher2006} that is commonly used in different fields of Astronomy to identify clusters without a previous knowledge of the sample \citep{Hareyama2019,Turner2019,Rubin2016}.
Given that we have seven color filters, each pixel of our terrain will have seven values corresponding to the reflectance of each band. In a seven-dimensional space (spectral space) each pixel is represented by a point of seven coordinates, and the totality of 
the pixels forms a point cloud. The clustering technique consists of identifying different subgroups within that cloud. If we assume that there are two representative subgroups, our partitioning algorithm K-Means generates two random points in our seven-dimensional space as representative clusters. Then, the algorithm calculates the distance of each point to each of the two representative clusters and, depending on the result, it assigns membership to one group or the other. After that, it calculates the centers of each data set, then computes all distances of each point to these new representative clusters, and 
iterates this process consecutively to refine the result. We show an illustration of this procedure in Fig. \ref{fig:kmeans}, where we have an image with four pixels measured with only three filters representing the three-dimensional cloud in spectral space. 

The result of this procedure is the identification of many representative clusters. We are interested not only in differences in reflectance, but also in the shape of the obtained spectra, as both are indicative of differences in composition. Therefore, we also apply the clustering analysis after normalizing the images. This clustering approach is more sensitive to shallow spectral features, such as the 700-nm absorption, as it excludes points with similar albedo but no 700-nm feature. To normalize the images, we divide the value of reflectance of each filter by the value of one of the filters that is used as a reference. To follow the commonly used procedure for ground-based photometric and spectroscopic data, we normalize the reflectance at 555 nm (filter F2). 

The most critical step of this procedure is the identification of the optimal number of clusters. Multiple statistical criteria can be applied to achieve this result, which, are not conclusive. Depending on the problem different approaches may be required. In this work we apply the Elbow method \citep{Syakur2018}, which allows us to find representative clusters by maximizing the difference between groups and the similarity between the members of each group. This method  computes the sum of squared errors (SSE) considering a range of the number of clusters and represents this quantity against the number of clusters. An elbow-like shape in this graphical representation points toward the suitable number. This elbow cannot always be unambiguously identified \citep{Ketchen1996}, 
especially in cases where there is a gradual and continuous transition of the data. In that case, the method gives a range of possible solutions that have to be inspected to determine the best one. This is a purely mathematical approach, and thus a subsequent physical interpretation of the results is always 
necessary.

To prepare our images, we first remove the pixels from the shadows of the craters, that is, areas not directly illuminated that could contaminate 
our results. We directly measure the reflectance in these shadowed regions and obtain an average reflectance of 0.065. Therefore, we eliminate all pixels with a 
reflectance lower than 0.065. As described in Section \ref{subsection2}, based on to the ranges where the selected photometric model performs suitably, we remove pixels with emission or incidence angles larger than 80$^{\circ}$ and phase angle values outside the range from 7$^{\circ}$ to 49$^{\circ}$. 
To reduce noise and eliminate possible unrepresentative pixels without loss of resolution, we use a mean 3x3 filter, replacing each pixel value in the image with the average value of its neighbors. To apply the K-means algorithm we use a code developed and provided by G. Marzo based on the description made in \cite{marzo2006} and using Interactive Data Language (IDL). The code slightly modifies the (now obsolete) IDL routine KMEANS to effectively manage more data than the original one. We added an additional modification in order to compute the sum of squared errors (SSE) parameter of the Elbow method.

\subsection{Characterization of spectral parameters} 
\label{subsection4}

In addition to identifying the number of clusters that describe our data set, we define several spectral parameters to characterize the obtained reflectance spectra. The first parameter is the visible spectral slope, $S$'. The slope is computed using the reflectance values at 555 nm (F2), 653 nm (F7), 749 nm (F3), 829 nm (F6), 917 nm (F4), and 965 nm (F5), normalized to unity at 555 nm (F2), by applying a linear fit using a least-squares method. Note that we do not use filter F8 (438 nm) to compute the spectral slope. This is mainly due to the fact that below 500 nm the reflectance spectra of most primitive, featureless asteroids suffer from a pronounced drop, due to the presence of an absorption feature centered to the ultra-violet. This drop-off in reflectance may produce values of spectral slope that are not representative of the overall slope of the spectra in the visible wavelength.

To compute the error in the spectral slope, we use a Monte Carlo approach \citep{Mario2018}, creating $1000$ clones in each of the filter bands by using a Gaussian distribution centered at the central wavelength of the filter and with a width corresponding to the uncertainty associated to that measurement. The resulting spectral slope is the average value in units of \%/1000 {\AA} and the error is the standard deviation. 

\begin{figure}[!htbp] 
\centering 
\includegraphics[width=\textwidth]{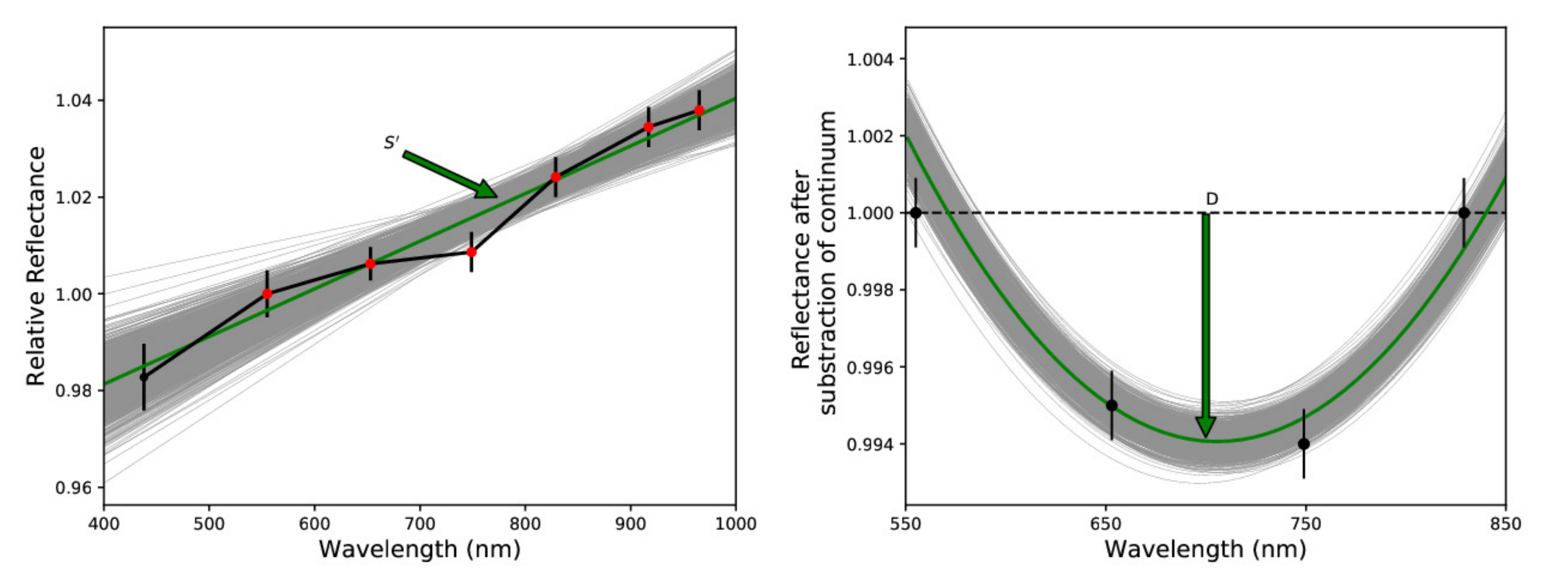} 
\caption{Example of the procedure to compute the spectral slope $S$' (left panel) and the band center and depth $D$ (right panel) of a relative reflectance spectrum, normalized to unity at 550 nm. The red points in the left panel are the reflectances used to compute the slope. The right panel zooms in on the absorption band for better visualization. The gray shadowed areas correspond to the 1000 clones generated using the Monte Carlo method and the final values of slope, band center, and band depth are computed as the average of these clones (green lines). In both cases, the error bars represent the standard deviation of the obtained average values.} 
\label{fig: slope} 
\end{figure}

If we detect an absorption band centered at 700 nm, the other spectral parameters are the wavelength position of the center of the band and the band depth, $D$. To measure these parameters, we first subtract the continuum between the 555 nm (F2) and 829 nm (F6) filters, then we fit a second-order polynomial to the  reflectances at 555 nm (F2), 653 nm (F7), 749 nm (F3), and 829 nm (F6). The wavelength position of the band center corresponds to the location at the minimum of the curve, whereas the band depth is computed as the difference, in percent, between a reflectance value of 1 and the reflectance value corresponding to the central wavelength position. The final values for the band center and depth and their corresponding errors are calculated using the same procedure as for the spectral slope. If the obtained value for the depth $D$ is above the 3$\sigma$ level, we consider this a positive detection of an absorption band. A graphical illustration of this procedure for both the slope and the band depth is shown in Fig. \ref{fig: slope}.


\section{Results} \label{Section4}

\begin{figure}[!htbp]
\centering 
\includegraphics[width=\textwidth]{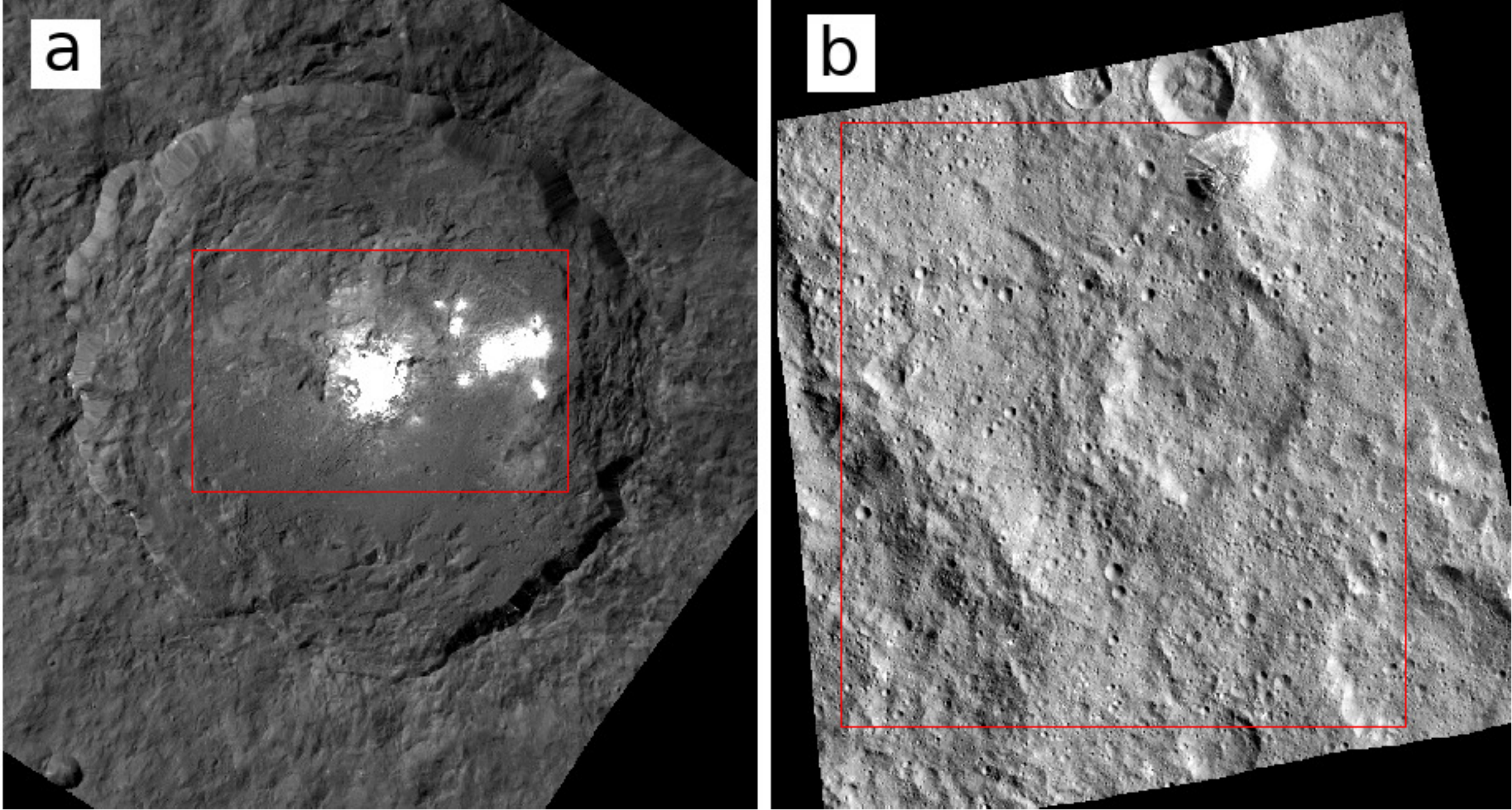} 
\caption{Images of the regions (red squares) that we analyzed: the interior of the Occator crater (a) and the Ahuna Mons and its surroundings (b).} 
\label{fig: areas} 
\end{figure}

We present the results obtained from applying the clustering technique to images of Occator crater and Ahuna Mons (Fig. \ref{fig: areas}, Table \ref{table: 1}). We calibrated the images in radiance factor, then projected and photometrically corrected them, as described above.

\subsection{Occator crater} \label{subsection5}

\begin{figure}[!htbp] 
\centering 
\includegraphics[width=8cm]{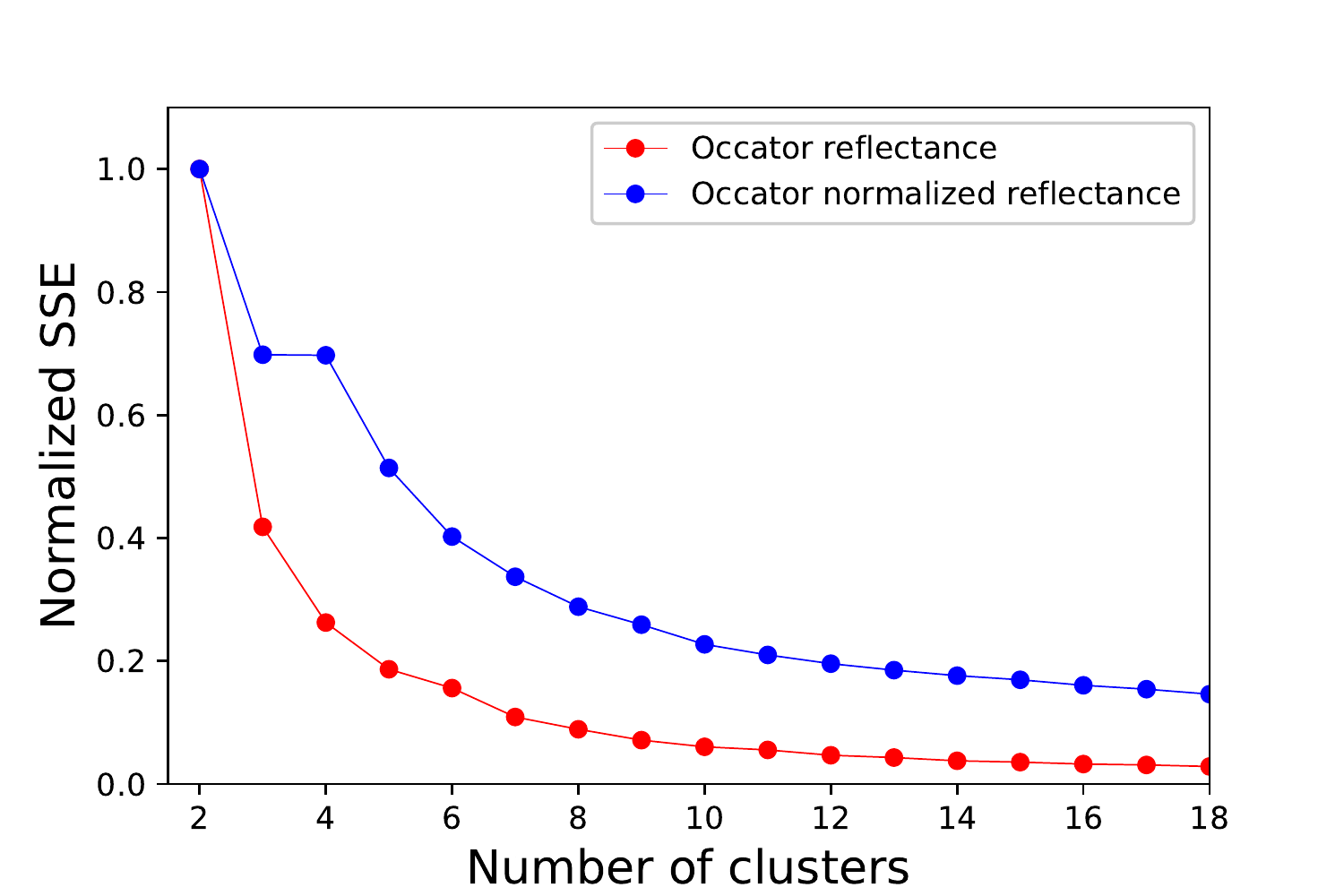} 
\caption{Normalized SSE {\it vs.} the number of clusters using the Elbow method for the Occator crater in reflectance (red) and normalized reflectance (blue). We see a  wide-open, shallow ``elbow'' for both reflectance and normalized reflectance cases. After analysis of all these possible solutions, we found an optimal number of seven clusters in reflectance and six in normalized reflectance.} 
\label{fig: elbow_occa} 
\end{figure}

\begin{figure}[!htbp] 
\centering 
\includegraphics[width=\textwidth]{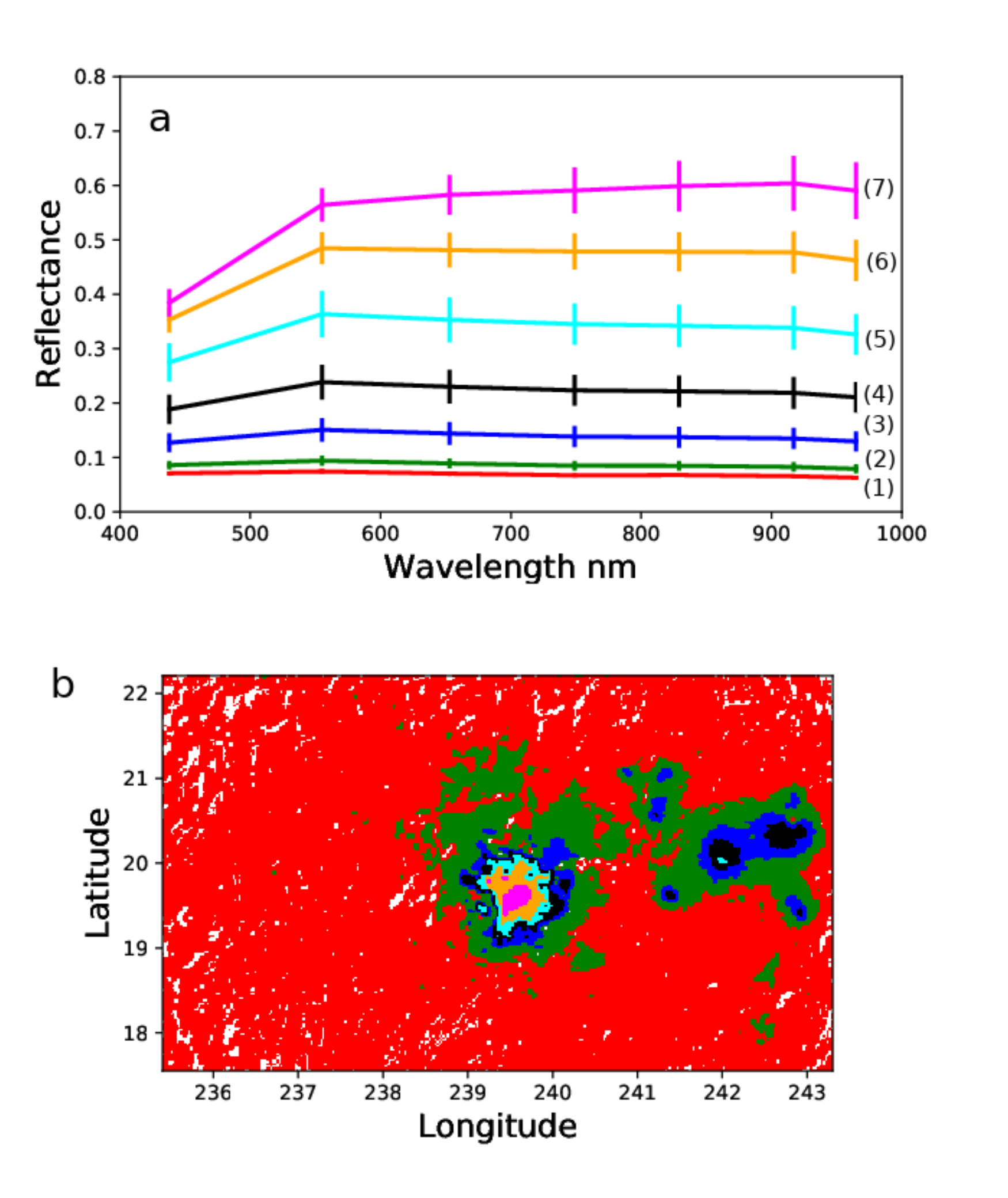} 
\caption{Results from the application of the clustering technique to the Occator crater. a) The seven representative spectral clusters with their corresponding error bars (standard deviation) are labeled with different colors. b) The location of the identified clusters over the analyzed area, using the same color code as in a). White pixels are those that are out of the defined limits: reflectance values lower than 0.065, emission and incidence angles larger than 80$^{\circ}$, and phase angles outside the 7$^{\circ}$ to 49$^{\circ}$ range.} 
\label{fig: 7clusters} 
\end{figure}

For this region, the Elbow method shows a shallow, wide-open elbow that extends up to nine/ten clusters, reaching an asymptotic regime (red points in 
Fig. \ref{fig: elbow_occa}). We consequently analyzed each possible solution in detail, studying how the pixels corresponding to each cluster were  distributed over the surface of the crater. We also investigated the spectral characteristics of each cluster to see when increasing the number of clusters produced specific regions with substantive spectral differences. We concluded that seven was the best choice to represent the orography of the surface and the continuous transition observed in albedo, from low values in the outer crater to high values in the inner regions. The resulting clusters are presented in Fig. \ref{fig: 7clusters} using different colors. For each cluster, we computed the spectral slope $S$', the band center, and the depth $D$. We also calculated the number of pixels contained in each cluster. The results are shown in Table \ref{table_occator_7} in increasing order of average reflectance. According to our detection criterion, we do not find any absorption band at 700 nm using this approach and taking into account albedo and spectral shape.

\begin{table}[!h] 
\small{ 
\begin{center} 
\begin{tabular}{lccccc} 
\multicolumn{5}{l}{\textbf{Occator Crater}}\\ 
\hline Cluster & $S$' (x10$^{-3}$) & $D$ & Center & $\bar{r}$ & Pixels \\ 
ID & (\%/1000 \AA) & (\%) & (nm) & (550 nm) & (\%)\\ 
\hline 
Red (1) & -3.3 $\pm$ 0.3 & 3.1 $\pm$ 1.3 & 698 $\pm$ 90 & 0.074 $\pm$ 0.004 & 81.4\\ 
Green (2) & -3.4 $\pm$ 0.6 & 3.0 $\pm$ 2.0 & 698 $\pm$ 17 & 0.093 $\pm$ 0.011 & 13.1\\ 
Blue (3) & -3.0 $\pm$ 1.4 & 3.2 $\pm$ 3.7 & 696 $\pm$ 27 & 0.149 $\pm$ 0.030 & 2.8\\ 
Black (4) & -2.5 $\pm$ 1.6 & 3.2 $\pm$ 3.9 & 696 $\pm$ 31 & 0.240 $\pm$ 0.052 & 1.3\\ 
Cyan (5)& -2.0 $\pm$ 1.7 & 2.8 $\pm$ 3.8 & 695 $\pm$ 26 & 0.375 $\pm$ 0.064 & 0.5\\ 
Orange (6) & -0.8 $\pm$ 1.2 & 1.8 $\pm$ 2.4 & 797 $\pm$ 26 & 0.490 $\pm$ 0.039 & 0.6\\ 
Fuchsia (7) & 1.4 $\pm$ 1.3 & 1.3 $\pm$ 2.2 & 706 $\pm$ 36 & 0.574 $\pm$ 0.043 & 0.3\\ 
\hline 
\end{tabular} 
\end{center} 
\caption{Computed spectral slope ($S$'), band depth ($D$), band center, and average reflectance at 550 nm ($\bar{r}$) for each of the seven representative clusters identified in the Occator crater (Fig. \ref{fig: 7clusters}). Errors correspond to the standard deviation. The last column shows the relative number of pixels 
(in \%) contained in each cluster. According to our defined criterion, there are no positive detections of the 700-nm hydration band.} \label{table_occator_7}} \end{table}

\begin{figure}[!htbp] 
\centering 
\includegraphics[width=\textwidth]{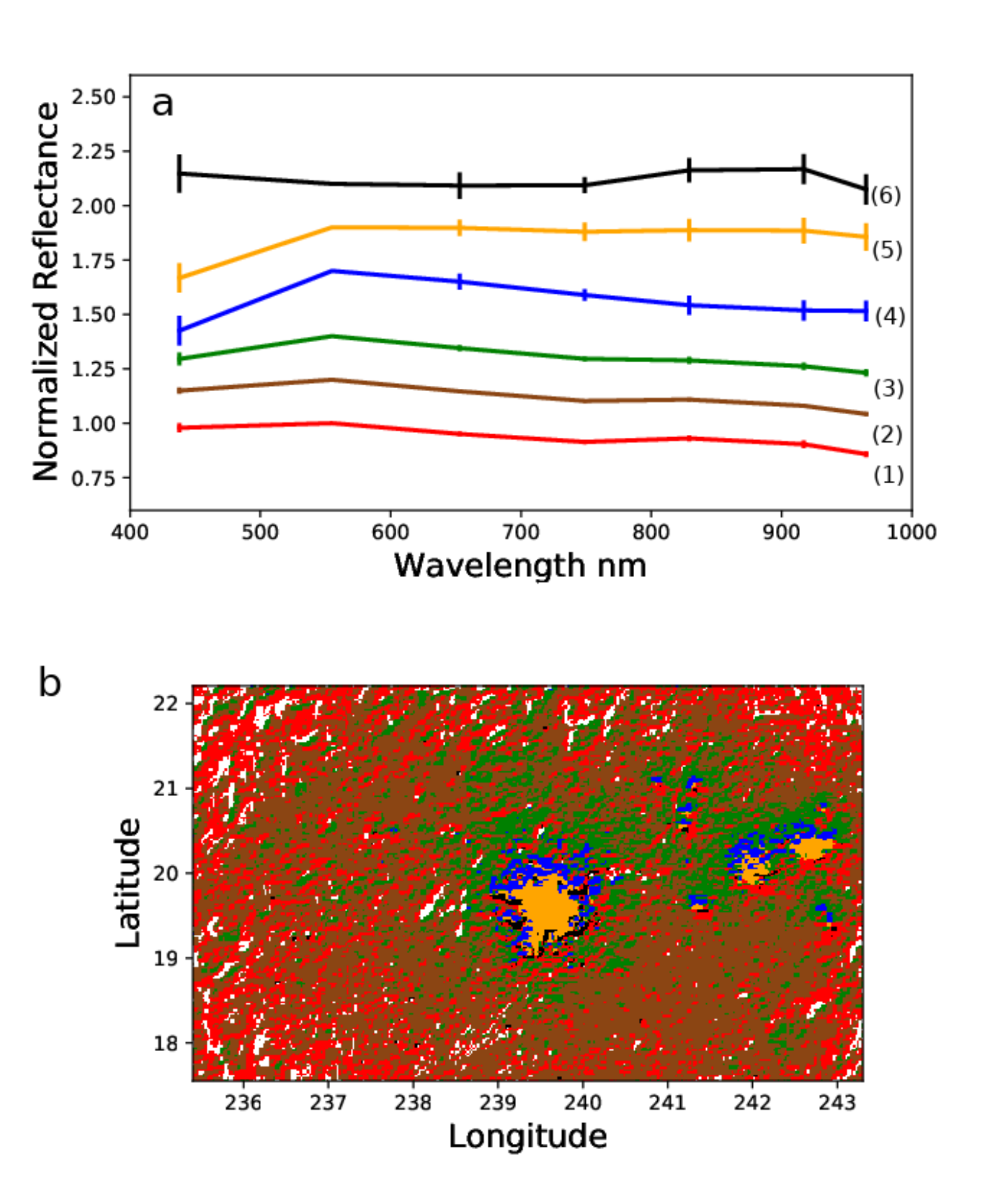} 
\caption{Results from the application of the spectral clustering to the Occator crater after normalization of the images at 555 nm. a) The six representative spectral clusters with their corresponding error bars (standard deviation) are labeled with different colors and numbers. The red and brown clusters, at the bottom of the plot, are the clusters where an absorption band at 700-nm is detected. b) The location of the identified clusters over the analyzed area, using the same color code as in a). White pixels are those that are out of the defined limits (as specified in Fig. \ref{fig: 7clusters}.)} 
\label{fig: norma_6_occator} 
\end{figure}

\begin{table}[!h] 
\small{ 
\begin{center} 
\begin{tabular}{lccccc} 
\multicolumn{5}{l}{\textbf{Occator Crater (normalized images)}}\\ 
\hline 
Cluster & $S$' (x10$^{-3}$) & $D$ & Center & $\bar{r}$ & Pixels\\ 
ID & (\%/1000 \AA) & (\%) & (nm) & (550 nm) & (\%)\\ 
\hline 
Red (1) & -2.8 $\pm$ 0.3 & 3.6 $\pm$ 1.1 & 698 $\pm$ 7 & 0.074 $\pm$ 0.015 & 23.9\\ 
Brown (2) & -3.3 $\pm$ 0.2 & 3.2 $\pm$ 0.9 & 698 $\pm$ 7 & 0.076 $\pm$ 0.007 & 56.8\\ 
Green (3) & -3.8 $\pm$ 0.4 & 2.4 $\pm$ 1.4 & 700 $\pm$ 18 & 0.092 $\pm$ 0.026 & 14.9\\ 
Blue (4) & -4.8 $\pm$ 1.0 & 1.3 $\pm$ 2.1 & 694 $\pm$ 33 & 0.206 $\pm$ 0.107 & 2.0\\
Orange (5) & -0.7 $\pm$ 1.2 & 2.1 $\pm$ 2.6 & 700 $\pm$ 32 & 0.379 $\pm$ 0.146 & 1.8\\ 
Black (6) & 0.8 $\pm$ 1.5 & 5.2 $\pm$ 4.3 & 698 $\pm$ 24 & 0.145 $\pm$ 0.077 & 0.6\\ 
\hline 
\end{tabular} 
\end{center} 
\caption{Same as Table \ref{table_occator_7} but for the six clusters identified in the normalized images of the Occator crater (Fig. \ref{fig: norma_6_occator}). In this case, $\bar{r}$ refers to the average reflectance at 550 nm computed before the normalization of the images. Only the red (1) and brown (2) clusters present band depth values above the 3$\sigma$ level and therefore are considered as positive detections of the 700-nm hydration band.} 
\label{table_occator}} 
\end{table}

We searched for the optimal number of clusters after normalizing the images, removing the albedo differences and emphasizing only the differences in spectral shape. In this case, the Elbow method showed also a wide elbow starting at four clusters and extending up to asymptotic regime (blue points in Fig. \ref{fig: elbow_occa}). After analyzing all possible solutions, we observed that when we increased the number of clusters from eight to nine, the code began to subdivide regions with spectral similarities in different clusters, where the new clusters were within the error bars of the old ones. Moreover, the associated pixels of these newly identified clusters were randomly dispersed in the images, producing a loss of any spatial coherence. Also, we identified two out of these eight clusters as outliers\footnote{We call outliers to those clusters that appear as a single isolated and ungrouped pixel, having no statistical significance.}. We therefore decided that six was the suitable number of clusters for the normalized images of the Occator crater (Fig. \ref{fig: norma_6_occator}).

As we did for the unnormalized images, we computed the spectral slope $S$', the band center, the band depth $D$, and the number of pixels for each cluster, and the resulting values are shown in Table \ref{table_occator}. In this case, the average reflectance at 550 nm ($\bar{r}$) corresponds to the average value of 
reflectance at 550 nm (F2) computed for the pixels of each identified cluster before the normalization process. In all cases the normalized spectra are flat, but considering the value of the average reflectance ($\bar{r}$) and spectral slope, we can form two different subgroups: one group containing the blue, orange, and black clusters with considerable spectral variations, and another group containing the first three clusters (red, brown, and green), all having negative slopes and low reflectances ($<$0.100). 

Within this group, two clusters (red and brown) present a positive detection of a band at 700 nm, having band centers at 698 $\pm$ 7 nm and band depths of 3.6 $\pm$ 1.1 and 3.2 $\pm$ 0.9 , respectively. Considering their spectral similarities, the properties of the 700-nm absorption band, the reflectance values at 550 nm, and the proximity of the pixels in the crater, we can combine them into a single cluster, presenting a band depth of 3.4 $\pm$ 1.0 . This value is in good agreement with that of 2.8 $\pm$ 1.2 obtained by \cite{Fornasier2014} for asteroids in the main belt, and with the values obtained by \cite{Morate2016,Morate2018} for asteroids in the Erigone and Sulamitis families. 

The 700-nm feature is still evident in the clusters that include albedo (Fig. \ref{fig: norma_6_occator}), but slightly below our threshold of 3$\sigma$. The spatial coherence of the normalized cluster with the 700-nm absorption (Fig. \ref{fig: blue}b) supports the validity and physical significance of this approach. 

It is important to remark here that we have performed several tests using a lower number of filters of the FC. As MapCam has only four filters, we did the exercise of repeating the above described procedure but using only four out of the seven filters of the Framing Camera: F8 (438 nm), F2 (555 nm), F7 (653 nm) or F3 (749 nm) and F6 (829 nm). These are the filters that are closer to those of MapCam: $b$' (473 nm), $v$ (550 nm), $w$ (698 nm) and $x$ (847 nm). The results obtained using a lower number of filters are basically the same. This applies also to the Ahuna Mons, and so it consists of an additional prove that the methodology can be applied to images of Bennu.

\subsection{Ahuna Mons} \label{subsection6}

\begin{figure}[!htbp] 
\centering 
\includegraphics[width=8.5cm]{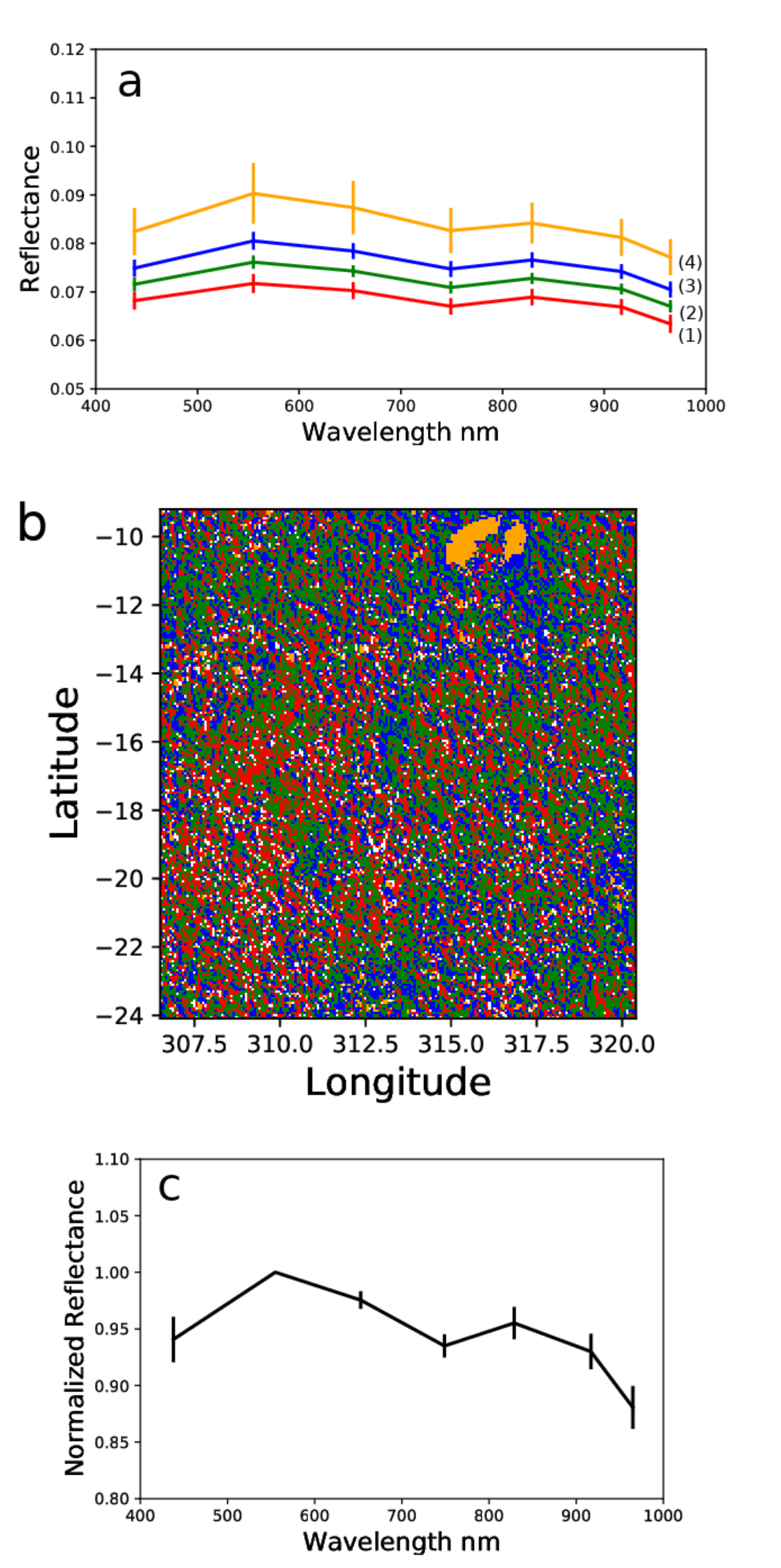} 
\caption{Results from the application of the clustering technique to the Ahuna Mons. a) The four spectral clusters and their corresponding error bars (standard deviation) are labeled with different colors. b) The location of the identified clusters over the analyzed area, using the same color code as in a). White pixels are those that are out of the defined limits (Fig. \ref{fig: 7clusters} ). c) Only one representative cluster is identified after the normalization of the images (error bars show standard deviation).} 
\label{fig: ahuna} 
\end{figure}

In this case, the Elbow method showed us a wide elbow from three to seven clusters, and we analyzed all the possible solutions within this range. We observed no considerable spectral differences, and each new possible solution was a subdivision of the reflectance variations due to noise---i.e., the fluctuations that we can expect after applying the photometric correction to the images. If we selected four clusters, they were scattered throughout the analyzed area, except the walls of the mons where one cluster concentrated in this region showed the highest reflectance (orange cluster in Fig.\ref{fig: ahuna}). The rest of the solutions presented the same behavior; the shapes of the clusters were very similar to each other, and the difference was just the number of subdivisions within the albedo fluctuations. When we ran the clustering code to the normalized images, the Elbow method lost coherence, indicating that all the pixels in the Ahuna Mons and its surrounding flat area could be represented by a single cluster (Fig. \ref{fig: ahuna}c). We computed the slope $S$', band center, and band depth $D$ for this single cluster, obtaining values of -1.4 $\pm$ 0.4 (x10$^{-3}$) \%/1000 \AA, 707 $\pm$ 10 nm, and 3.2 $\pm$ 1.1 \%, respectively. The computed band depth was just below the 3$\sigma$ level, and therefore we had no positive detections of the 700-nm hydration band in the Ahuna Mons. 

\section{Discussion} 
\label{Section5}

\subsection{Occator crater}

This crater (${19.82}^{\circ}$ N, ${239.33}^{\circ}$ E) is the highest albedo feature among the more than $150$ on Ceres. With a diameter of $90.5$ km and a depth of $4$ km, it has a large central pit $10$ km wide that is traversed by dark lineaments. Using the Dawn Visible and Infrared Mapping Spectrometer, 
\cite{DeSanctis2016} found that bright material inside the Occator crater is abundant in salts such as anhydrous sodium carbonate (Na$_2$CO$_3$) and ammonium salts. This composition is different from the rest of the surface of Ceres. They also found that these high-albedo materials are incompatible with impact 
excavation and that the salts suffered a percolated process toward the surface. Near-infrared observations \citep{Ruesch2016,Zambon2017} identified minor phyllosilicates and opaque minerals in addition to anhydrous sodium carbonate, also detected at the cryovolcanic dome of the Ahuna Mons. Occator is a young crater, about 100 million years old, and sublimation of water ice has been suggested by \cite{Nathues2015} as a possible explanation for the bright materials inside the crater. According to \cite{DeSanctis2016}, \cite{Zolotov2017} and \cite{Vu2017} a brine consisting of liquid water with an abundance of dissolved salts (mostly carbonates), gases, and entrained altered solids would be in the subsurface. 

\begin{figure}[!htbp] 
\centering 
\includegraphics[width=\textwidth]{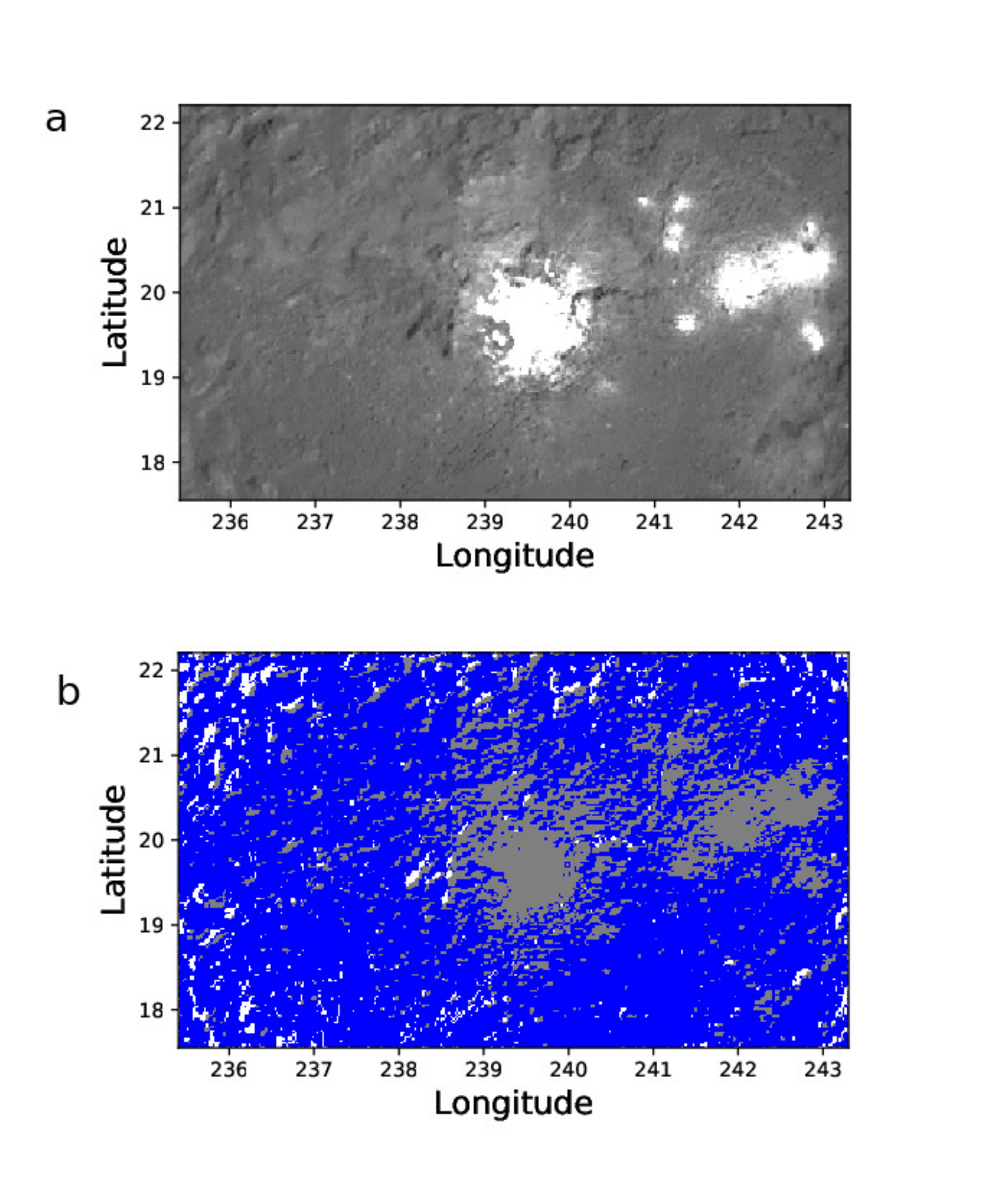} 
\caption{Region inside the Occator crater analyzed in this work. a) Calibrated image. b) Image pixels corresponding to the clusters where the 700-nm band is detected are shown in blue, whereas those corresponding to no detections are shown in grey. White pixels are those that are out of the defined limits (Fig. \ref{fig: 7clusters}).} 
\label{fig: blue} 
\end{figure}

In our first approach, the spectral clustering analysis showed us seven clusters (Fig. \ref{fig: 7clusters}), distributed from the outer regions of the crater with lower reflectance (albedo) to the central part where the bright spots are located. The clusters have a distribution in spectral slope, with the outer regions having more negative slopes. Although none of the computed band depths comply with the 3$\sigma$ criterion, they increase toward the outer parts of the crater. This apparent connection between the shape of the spectra and the albedo is even more evident when we normalize the images. Our clustering code grouped the 
normalized spectra into six clusters (Fig. \ref{fig: norma_6_occator}). The lower albedo region is divided into two subregions: one area with considerable detection of the absorption band at 700 nm, indicating hydrated minerals (blue pixels in Fig. \ref{fig: blue}b), and another area where this band is not detected (gray pixels in Fig. \ref{fig: blue}b). In both cases, the clusters correspond to the orography. 

The two clusters showing the 700-nm hydration absorption band (which we combined into one) are located in the most distant regions from the Vinalia faculae and Cerealia facula, which are bright and salt-rich. The faculae's proximity to each other and their shared morphological and compositional properties supports 
their similar formation mechanisms as claimed by \cite{Ruesch2019}: the best hypothesis for their formation is the transport of liquid brine, reaching the surface as salt-water fountains. The detection of the 700-nm absorption band in the outer regions of the Occator crater reinforces the hypothesis of a liquid brine lodged in the subsurface that, through this transport mechanism, is altering the surrounding minerals. Unlike in the central parts where the salts dominate, the interaction of 
the silicates with the water in the outer regions provides a way to form the hydrated minerals. 

We found a strange behavior in one of the six clusters identified in the normalized images of the Occator crater. The black cluster in Fig. \ref{fig: norma_6_occator} contains only 0.6 \% of the pixels, concentrated in the transitional zone between the low-albedo pixels and the bright spots. This cluster shows a normalized reflectance at 438 nm that is considerably higher than that of the rest of the clusters and a noticeable drop in reflectance in the 965-nm band. Given that it is located in this transitional zone, a possible explanation is that this is an orographically irregular area with variety of slopes and shadows and therefore the photometric corrections, given the limited resolution of our shape model, are not reliable.

\subsection{Ahuna Mons}

Ahuna Mons (${10.46}^{\circ}$ S, ${315.8}^{\circ}$ E) is the largest mountain on Ceres, with an average height of about $4$ km. This particular area is of interest because of recently discovered changes in the presence of hydrated sodium carbonates \citep{Carrozzo2018}, indicating that Ceres is a geologically and 
chemically active body. This mountain is interpreted as a geologically young cryo-volcanic dome \citep{Sori2017}---i.e., a volcano that erupts water, ammonia or other volatile materials rather than magma. Owing to its distinctive characteristics, we selected this area to apply our method and search for spectral 
signatures. 

Ahuna Mons and its surroundings show four similar clusters scattered throughout the area. All four clusters yielded by our algorithm have the same spectral shape, and are simply a representation of a continuum in albedo that fluctuates around an average value. When we apply the clustering code to the normalized images, we only find one representative cluster of the entire surface (Fig. \ref{fig: ahuna}c). 

The top of the Ahuna Mons has precisely the same spectral behavior as its surroundings. This finding supports the suggestion that cryo-vulcanism created the mountain and covered the surrounding terrain. 

\section{Summary}

We present here a methodology based on spectral clustering techniques. We applied it to Ceres, using color images in seven filters obtained with the Framing Camera onboard NASA's Dawn spacecraft. The OSIRIS-REx spacecraft obtains color images of Bennu with MapCam, visible camera equipped with four color 
filters spanning a similar wavelength range to that of the Framing Camera. In addition, the two missions follow similar observational strategies, acquiring images at low phase angles. Furthermore, both Ceres and Bennu are primitive asteroids, showing low-albedo, almost featureless spectra. Thus, we validate our clustering methodology using images of Ceres in preparation for its application to MapCam images of Bennu.

We selected two particularly interesting areas on the surface of Ceres: the Occator crater, where bright spots suggest the existence of salts, and the Ahuna Mons, the largest mountain on the dwarf planet, where hydrated sodium carbonates recently have been discovered. We downloaded the images of these two regions 
from the PDS and projected them using the ISIS3 software. Calibrated images from the PDS presented inconsistencies with data in the literature, so we downloaded raw images and performed the calibration using the procedure described by \citet{Schroder2013,2014Schro}. The final step before applying our spectral analysis was to photometrically correct the images using the parameter-less Akimov model after calculating all the necessary coefficients for each color filter. 

We applied our clustering technique based on a K-means algorithm together with the Elbow method. This technique allows us to identify regions with spectral similarities, which is useful for analyzing the enormous amount of information that will come from MapCam. 

For the Occator crater, we found a robust correlation between the spectral clusters and the terrain morphology. This apparent relationship among the orography, the albedo values, and the spectra was evident in both reflectance and normalized reflectance. According to our 3$\sigma$ criterion, we detected the 700-nm absorption band, associated with hydrated silicates, in one of the normalized reflectance clusters, with a depth of 3.4 $\pm$ 1.0 \%. When spectra are normalized before analysis, the clustering approach is more sensitive to shallow spectral features, such as the 700-nm absorption, because it excludes points with 
similar albedo but no 700-nm feature. The pixels corresponding to this cluster were located in the outer regions of the Occator crater floor. This absorption is furthest away from the central bright spots of the crater, which is consistent with their liquid-brine origin suggested by \cite{Ruesch2019}. 

However, we found no evidence of hydrated minerals or any other unusual spectral characteristic in the Ahuna Mons region. The top of Ahuna Mons has the same spectral behavior as the rest of its surroundings, consistent with its cryo-volcanic origin suggested by \cite{Sori2017}.

\section*{Acknowledgments}

The authors thank Stefan Schr{\"o}der for his advice and assistance with the calibration procedure and Song Seto for her 
useful suggestions and comments. J. L. Rizos, J. Licandro, J. de Le\'on and M. Popescu acknowledge support from the AYA2015-67772-R (MINECO, Spain) J. de Le\'on acknowledges financial support from the Severo Ochoa Program SEV-2015-0548 (MINECO) and the project ProID2017010112 under the Operational Programmes of the European Regional Development Fund and the European Social Fund of the Canary Islands (OP-ERDF-ESF), as well as the Canarian Agency for Research, Innovation and Information Society (ACIISI). H. Campins acknowledges support as a Co-Investigator on NASA's OSIRIS-REx 
mission under Contract NNM10AA11C issued through the New Frontiers Program.


\bibliographystyle{elsarticle-harv}

\bibliography{references}

\end{document}